\documentclass{article}

\usepackage{amsmath}
\usepackage{amssymb}
\usepackage{mathptmx}       
\usepackage{helvet}         
\usepackage{courier}        
\usepackage{type1cm}        
\usepackage{makeidx}         
\usepackage{graphicx}        
\usepackage{multicol}        
\usepackage[bottom]{footmisc}
\usepackage{listings}
\usepackage{color}
\definecolor{ForestGreen}{rgb}{0.13,0.54,0.13}

\newtheorem{definition}{Definition}

\newtheorem{proposition}{Proposition}

 \newcommand {\be}{\begin{equation}}
\newcommand {\ee}{\end{equation}}
\begin {document}

\title{Formal assessment of some properties of Context-Aware Systems}


\author{Fabio A. Schreiber\footnote{Fabio A. Schreiber was partially supported by INAIL, RECKON Project.
} \\
Dipartimento di Elettronica, Informazione e Bioingegneria\\
Politecnico di Milano, 
Milan, Italy\\
\and 
Maria Elena Valcher\\
Department of Information Engineering\\
University of Padova 
Padova, Italy}
\maketitle

{\small
{\bf Abstract.}
Context-Aware systems are becoming useful components in autonomic and monitoring applications and the   assessment of their properties is an important step towards reliable implementation,  especially in safety-critical applications. In this paper, using an avalanche/landslide alert system as a running example, we propose a technique, based on Boolean Control Networks, to verify that the system dynamics has  stable equilibrium states, corresponding to constant inputs, and hence it does not exhibit   oscillatory behaviors, and to establish other useful properties in order to implement a precise and timely alarm system. 


}





\setcounter{page}{1}

\section{Introduction}\label{intro}

In the first days of 2017 a hotel has been hit by an avalanche in Abruzzo, an Italian region, causing the death of 29 people. Preliminary technical
findings stated that the incident was triggered by a series of earthquakes in central Italy, in conjunction with the raise of the atmospheric
temperature that melted the snow. In such cases, when the alarm is spread, the security manager is confronted with two alternatives: i) evacuate the site, which often is a cumbersome and uncomfortable operation which spreads discontent among the guests if it results in a false alarm; ii) do nothing, in the hope that nothing dangerous will happen.  In this paper, we address the problem of  formalizing,  by means of a context-Aware (C-A) system, a decision process to avoid this type of tragedies as well as false alarms. 
In addition, by making use of the algebraic approach to Boolean Control Networks, we are able to 
assess the   existence of globally attractive equilibrium points of  the overall decision system, corresponding to constant inputs, and to investigate some interesting structural properties, that  formalize system features of great practical relevance. In detail,  in Section \ref{related} we introduce the background upon which our research is founded; in Section \ref{architecture} we describe the system architecture and in Section \ref{BCN} we briefly introduce the Boolean Control Networks algebraic description.  Section \ref{model} describes the BCN model for the hydrogeological example. In Section \ref{equilibria}  it is shown that the BCN exhibits only globally attractive equilibrium points, and no                                                                                                                                                                                                                                                                                                                                                                                                            
 limit cycles, corresponding to constant inputs; in Section \ref{observability} the observability and reconstructibility of the system are considered; Section \ref{fault} examines the possibility of identifying some kinds of faults in the inputs that could result in errors in the alarm system,  and, in  Section \ref{conclusion},  some conclusive remarks and hints for future work are made.

\section{Related works}\label{related}

Context-aware computing was born out of the need to master the complexity and enhance the flexibility of modern computing and information systems.  Among the most widely used definitions of Context, and of Context-aware Computing, those proposed by A. Dey \cite{Dey2001} state: {\sl{``Context is any information that can be used to characterize the situation of an entity. An entity is a person, place, or object that is considered relevant to the interaction between a user and an application, including the user and applications themselves."}} and {\sl{``A system is Context-aware if it uses context to provide relevant information and/or services to the user, where relevancy depends on the user’s task."}} \\ 
Accordingly, sophisticated and general Context models have been proposed, to support Context-aware applications that use them to:  (i) tailor the set of application-relevant data, (ii) increase the precision of information retrieval, (iii) discover services, (iv) build smart environments, and others \cite{Bolchini2007}. \\
In the domains of Databases and Programming Languages, the design of Context-aware and Self-adapting systems has been frequently based on  the separation between Context and functional system \cite{Bolchini2009,Schreiber2017}.  Even if a holistic view of the Context-aware system in which the Context and the functional system variables sets are kept together is possible, a separation of the two sets has been advocated, by using a component-based approach, to master the growing complexity of modern software systems and enforcing the separation of concerns \cite{Djoudi2016}.\\
The introduction of Context-awareness and Self-adaptation in safety-critical applications arose the need of specifying and assessing their properties, mainly those related to the system dependability, by means of formal methods such as Bigraphs and model-checking \cite{Djoudi2016,Cherfia2014}. Owing to the dynamic nature of self-adapting systems,   \emph{stability} has drawn great attention among the features affecting dependability. Nzekwa et  Al. \cite{Nzekwa2010} propose the composition of different mechanisms to obtain a flexible model for implementing stabilization in Context-aware systems.\\
In \cite{Padovitz2004,Padovitz2005} Padovitz et Al. consider a state-space approach for describing the \emph{situation} dimension and for determining the likelihood of transitions between  \emph{situation subspaces}, all other Context dimensions remaining constant. In their model, the state variables are constituted by the system's sensors outputs. In  an analog system, many sets of sensors values, representing a system state, 
can belong to the same situation subspace as far as they satisfy the conditions in its defining expression; a transition starts whenever one or more values change in such a way as to respectively switch the expressions for two situation subspaces from \texttt{TRUE} to \texttt{FALSE} and vice-versa. The likelihood of the transition is evaluated by assuming notions analogous to those of velocity and acceleration in mechanical systems, and on the basis of the distance of the values of the actual situation from those of its boundary.
\\
Stability is a traditional topic in control systems theory, and in \cite{Diao2005} the authors explore {\sl{``... the extent to which control theory can provide an architectural and analytic foundation for building self-managing systems ..."}}. However, control systems are typically described by means of differential equations and by Matrix Algebra, while Context-aware systems are   digital
and mostly based on Logics. Inspired by biological systems, Boolean Networks (BN) and Boolean Control Networks (BCN) have been introduced, their representative equations have been converted into an equivalent algebraic form \cite{Cheng2010b,Cheng2010a}, and solutions to problems such as controllability, observability, stability and reconstructibility have been proposed \cite{Cheng2009,EF_MEV_BCN_obs2012,Fova2016}. \\
We think that cooperation between the two disciplines can be fruitful, therefore, to fill the model gap, in this paper we model the Context as well as the functional system as  Boolean Control Networks, as briefly explained in Section \ref{model}.

\section{The architecture of the monitoring system}\label{architecture}

\begin{figure}[h]
	\vspace{-2.5 cm}
		\includegraphics[scale=0.65]{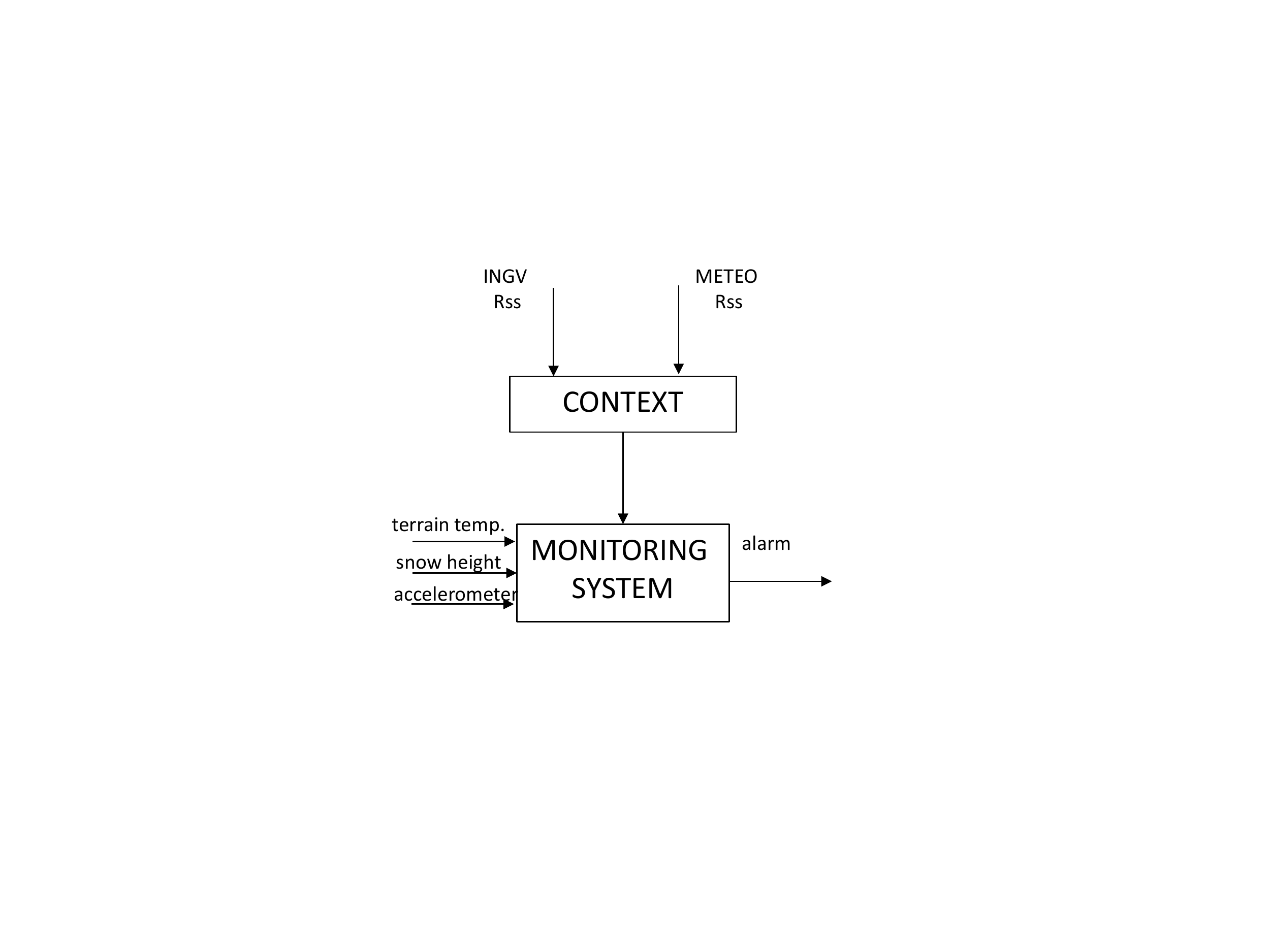}
	\vspace{-4.2cm}
	\caption{Open loop Context-Aware system}
	\label{fig:C-A system}
\end{figure}

Figure \ref{fig:C-A system} shows the general architecture of a Context-aware system \cite{Bolchini2007,Dey2001} conceived for monitoring possible snow/ground slides.  Signals, coming from physical sensors on the ground, are evaluated in the context of the seismic and meteorological information provided by Web Services RSSs -  which can suggest immediate danger - in order to issue alarms. Combining the Context state with the actual physical data that are input to the functional system allows the design of a flexible and effective prevention information system which, as an example, can distinguish between the vibration caused by the detachment of a snow mass  and that caused by a skier or a deer occasionally passing near a sensor.  

In the monitoring system 
 some states produce outputs that can affect the environment,   e.g. by possibly activating an alarm siren. 
  In case of an alarm, the time to evacuate a hotel can be in the order of hours, while the seismic and meteorological conditions can change faster.  The ultimate goal of this study is to be sure that in dangerous situations an alarm signal is issued, but at the same time  that frequent changes in the Context State do not  induce an oscillatory behavior of the alarm system and the consequent movement of people out and back into the hotel. The designer of the C-A system must ensure that no action is started before the preceding one is terminated (e.g. reducing the evacuation time). In this paper we use a simple open-loop model; however, in    more complex C-A self managing applications, the system output can affect the context itself.
 
Our aim is therefore:

\begin{itemize}
\item To describe a Context-Aware (C-A) system, as in Figure \ref{fig:C-A system}, by means of a logic State Space model;  web services provide input messages to the Context and sensors provide input signals to the Monitoring System; the Context state is a further input to the latter.
\item To use BCNs and System Theory tools to asses the properties of a C-A system: the existence of globally stable equilibrium points and the absence of oscillatory behaviors (limit cycles) under constant inputs; the reconstructibility of the system and the detection of some faults affecting the C-A system inputs.
\end{itemize}

\section{Algebraic representation of Boolean Control Networks}\label{BCN}

We consider Boolean vectors and matrices, taking values in ${\mathcal B}  = \{0,1\}$, with the usual
logical operations (And $\wedge$, Or $\vee$,   Negation  $\neg$). $\delta^i_{k}$   denotes the $i$th canonical vector of size $k$, 
namely the $i$th column of the $k$-dimensional identity matrix $I_k$. $\Delta_{k}$ is
the set of  all  $k$-dimensional canonical vectors, and ${\mathcal L}_{k\times n}\subset {\mathcal B}^{k\times n}$  the set of all $k\times n$ {\em logical matrices}, namely $k\times n$ matrices whose  columns are   canonical vectors of size $k$. 
Any matrix $L\in  {\mathcal L}_{k\times n}$ can be represented as 
a row vector whose entries are canonical vectors in ${\mathcal L}_{k}$, namely
$L= \left[\begin{matrix}\delta_k^{i_1} &\delta_k^{i_2} &\dots & \delta_k^{i_n}\end{matrix}\right],$ for suitable indices $i_1,i_2, \dots, i_n\in [1,k].$
$[A]_{\ell j}$ is the
  $(\ell,j)$th entry of the matrix $A$.
 A 
nonsingular square matrix in ${\mathcal L}_{k \times k}$ is a {\em permutation matrix}.

There is a bijective correspondence between
 Boolean variables $X\in {\mathcal B}$ and vectors ${\bf x}\in \Delta_2$, defined by the relationship
 $${\bf x} = \left[\begin{matrix} X\cr \overline{X}\end{matrix}\right].$$
We introduce the {\em (left) semi-tensor product} $\ltimes$ between matrices (and hence, in particular, vectors) as follows \cite{BCNCheng}:
given $L_1\in {\mathcal L}_{r_1 \times c_1}$ and $L_2\in {\mathcal L}_{r_2\times c_2}$, we set 
$$L_1\ltimes L_2 := (L_1 \otimes I_{T/c_1})(L_2 \otimes I_{T/r_2}),
\quad {\rm where}\quad T:= {\rm l.c.m.}\{c_1,r_2\}.$$
The semi-tensor product represents an extension of the standard matrix product, by this meaning that if $c_1=r_2$, then
$L_1 \ltimes L_2=L_1L_2$.
Note that if ${\bf x}_1\in \Delta_{r_1}$ and ${\bf x}_2\in \Delta_{r_2}$, then
${\bf x}_1 \ltimes {\bf x}_2\in  \Delta_{r_1r_2}.$
 By resorting to the semi-tensor product, we can extend the previous   correspondence to a bijective correspondence   \cite{BCNCheng} between ${\mathcal B}^n$ and ${\mathcal L}_{2^n}$. This is possible in the following way: given $X= \left[\begin{matrix}X_1 & X_2 & \dots & X_n\end{matrix}\right]^\top\in {\mathcal B}^n$ 
set  
$${\bf x} := \left[\begin{matrix}X_1\cr \overline{X}_1\end{matrix}\right] \ltimes \left[\begin{matrix}X_2\cr \overline{X}_2\end{matrix}\right]\ltimes \dots \ltimes \left[\begin{matrix}X_n\cr \overline{X}_n\end{matrix}\right].$$
This  corresponds to
$${\bf x}= \left[\begin{matrix}X_1X_2\dots X_{n-1} X_n \cr X_1X_2\dots X_{n-1} \ \overline{X}_n \cr  X_1X_2\dots \overline{X}_{n-1} X_n \cr \vdots \cr
\overline{X}_1\overline{X}_2\dots \overline{X}_{n-1} \overline{X}_n\end{matrix}\right].$$
A {\em Boolean Control Network}  (BCN) is  a logic state-space model taking the following expression:
\be
\begin{array}{rcl}
X(t+1) &=& f(X(t),U(t)), \cr
Y(t)&=& h(X(t),U(t)), \qquad t \in \mathbb{Z}_+,
\end{array}
\label{BCNL}
\ee
where $X(t)$, $U(t)$  and $Y(t)$ are
 the $n$-dimensional state variable,  the $m$-dimensional input variable and the $p$-dimensional output variable at   time $t$, taking values in ${\mathcal B}^n$,    ${\mathcal B}^m$ and ${\mathcal B}^p$, respectively.
 $f$ and $h$ are  logic functions, i.e. $f: {\mathcal B}^n \times {\mathcal B}^m \rightarrow {\mathcal B}^n$, while 
 $h: {\mathcal B}^n \times {\mathcal B}^m \rightarrow {\mathcal B}^p$.
By resorting to the   semi-tensor product $\ltimes$, the BCN (\ref{BCNL}) can be described as \cite{BCNCheng}
\be
\begin{array}{rcl}
{\bf x}(t+1)  &=& L \ltimes {\bf u}(t)\ltimes {\bf x}(t), \cr
{\bf y}(t) &=& H \ltimes {\bf u}(t) \ltimes {\bf x}(t), \qquad t \in \mathbb{Z}_+,
\end{array}
\label{BCNA}
\ee
where $L \in \mathcal{L}_{2^n \times 2^{n+m}}$ and $H \in \mathcal{L}_{2^p \times 2^{n+m}}$. This is called the {\em algebraic expression} of the BCN. The matrix $L$ can be partitioned into $2^m$ square blocks of size $2^n$, namely as
$$L = \begin{bmatrix} L_1 & L_2 & \dots & L_{2^m}\end{bmatrix}.$$
For every $i\in \{1,2,\dots, 2^m\}$, the matrix $L_i\in {\mathcal L}_{2^n\times 2^n}$ represents the logic matrix that relates ${\bf x}(t+1)$ to ${\bf x}(t)$, when ${\bf u}(t)=\delta^i_{2^n}$, namely
$${\bf u}(t)=\delta^i_{2^m} \ \Rightarrow \ {\bf x}(t+1)= L_i {\bf x}(t).$$
It is worth remarking that the previous algebraic expression \eqref{BCNA} can be adopted to represent any state-space model in which the state and input variables take values in finite sets, and hence the sizes of the state and input vectors  need  not   be powers of $2$. In that case  oftentimes BCNs are called multi-valued Control Networks \cite{BCNCheng}. With an abuse of terminology, in this paper we will always refer to them  as BCNs. Moreover, we will replace $2^n, 2^m$ and $2^p$ with the generic symbols $N, M$ and $P$.

\section{The BCN Model of the Hydrogeological Example}\label{model}

\subsection{The Context model}\label{context}

\subsubsection{Context Input Variables}

The values of the Context Input Variables are supplied by RSS messages coming from National Web Services, such as Meteorological forecasts and the National Geophysics Institute. Even if the message frequency can be variable, for ease of modeling, we suppose that the system samples them with the same frequency. Moreover, we suppose that a real danger situation can be expected only when a defined number - in our example at least four - of consecutive earthquake announcements are sent together with a snow forecast.\\ 
We assume:

INGV\{$earthquake, \neg earthquake$\} := U$_{1}$

METEO\{$snow, \neg snow$\} :=U$_{2}$
\\
Therefore,  by  expressing the context input variables in terms of canonical vectors, we get:\\ \\
\textbf{Context Input Vector}:\\

  {\bf u}(t)=
$\left|\begin{array}{cc}
U_{1}&U_{2}\\U_{1}&\neg U_{2}\\\neg U_{1}&U_{2}\\\neg U_{1}&\neg U_{2}
\end{array}\right|
\in \Delta_4:= \{\delta^{1}_{4},\delta^{2}_{4},\delta^{3}_{4},\delta^{4}_{4}\}$
\\ 

\subsubsection{Context States} 

 As previously mentioned, we assume that simultaneous snow and earthquake alerts can be regarded as reliable only if not isolated, namely if a sufficiently high number of consecutive 
(simultaneous) alerts are sent (and received). For this reason we introduce as Context State a counter:\\ \\
COUNTER\{$0,1,2,3,>3$\}  =: C
\\ \\
In the representation by means of canonical vectors, the counter   is denoted by ${\bf c}$ and  takes values in $\Delta_5 := \{\delta^1_{5},\delta^2_5,\delta^3_5,\delta^4_5,\delta_5^5\}$, depending on how many consecutive simultaneous alerts for snow and earthquake have been received.
Specifically, for $i=1,2,3,4$, we have that ${\bf c}(t)=\delta^i_5$ if the counter is $i-1$ at time $t$, while
${\bf c}(t)=\delta^5_5$ if the counter is at least $4$ at time $t$.\\
If the counter at time $t$ has a value in  $\{\delta^1_{5},\delta^2_5,\delta^3_5,\delta^4_5\}$ and the context input is ${\bf u}(t)=\delta^1_4$ (another simultaneous snow and earthquake alert comes in), then the counter value at $t+1$ is increased by $1$. If ${\bf c}(t)=\delta^5_5$ and ${\bf u}(t)=\delta^1_4$, then 
${\bf c}(t+1)=\delta^5_5$, while in every other case the counter is reset\footnote{This is one  possible solution, but it may be regarded as somewhat dangerous:   if the counter gets erroneously reset, then  the alert ends up being significantly delayed. An alternative solution could be that of simply decreasing by one the counter if ${\bf u}(t) \ne\delta^1_4$ (or if ${\bf u}(t) =\delta^i_4, i=2,3$). This solution would be more robust to possible disturbances occasionally affecting the context inputs.} to ${\bf c}(t+1)=\delta^1_5$.\\
Therefore, the counter updates according to the following model (BCN):\\
\noindent ${\bf{c}}(t+1) 
= {\bf{C}}\ltimes {\bf{u}}(t)\ltimes {\bf{c}}(t)$, where\\

\noindent $C = \begin{bmatrix} C_1 & C_2 & C_3 & C_4\end{bmatrix}\in {\mathcal L}_{5\times 20}$, and 
\\

\noindent $C_{1}= C \ltimes \delta^{1}_{4} = [\delta^{2}_{5}\ \delta^{3}_{5}\ \delta^{4}_{5}\   \delta^{5}_{5} \  \delta^{5}_{5}]$\\
$C_{2}= C \ltimes \delta^{2}_{4} = [\delta^{1}_{5}\ \delta^{1}_{5}\ \delta^{1}_{5}\   \delta^{1}_{5}\ \delta^1_5]$\\
$C_{3}= C \ltimes \delta^{3}_{4} =C_2$\\
$C_{4}= C \ltimes \delta^{4}_{4} = C_2$\\

Obviously, \emph{the number of consecutive alert situations is a design variable which allows to set more stringent - if increased - or more relaxed - if lowered - requirements on the alarm system.} \\ 


\subsubsection{Context Output}


 Introduce the Context model output\\

{CONTEXT-ALERT\{$danger, quiet$\} := U$_c$}\\

 We assume that the CONTEXT-ALERT variable U$_c$  
 is \emph{danger} (the corresponding canonical vector {\bf u}$_c$ takes the value $\delta^1_2$) if and only if there have been at least four simultaneous snow and earthquake alerts.
\\
So, the variable  ${\bf u}_c$ updates following the   algebraic rule:\\
\noindent ${\bf{u}}_c(t) ={\bf H_c}  \ltimes {\bf{c}}(t)$
\\
where \\ 
${\bf H_c}$
= $ \left[ \delta^{2}_{2} \   \delta^{2}_{2} \   \delta^{2}_{2} \  \delta^{2}_{2} \ \delta^1_2 \right] \in {\mathcal L}_{2\times 5}$   
\\

\begin{figure}[h]
\vspace{-1.0 cm}
	\includegraphics[scale=0.4]{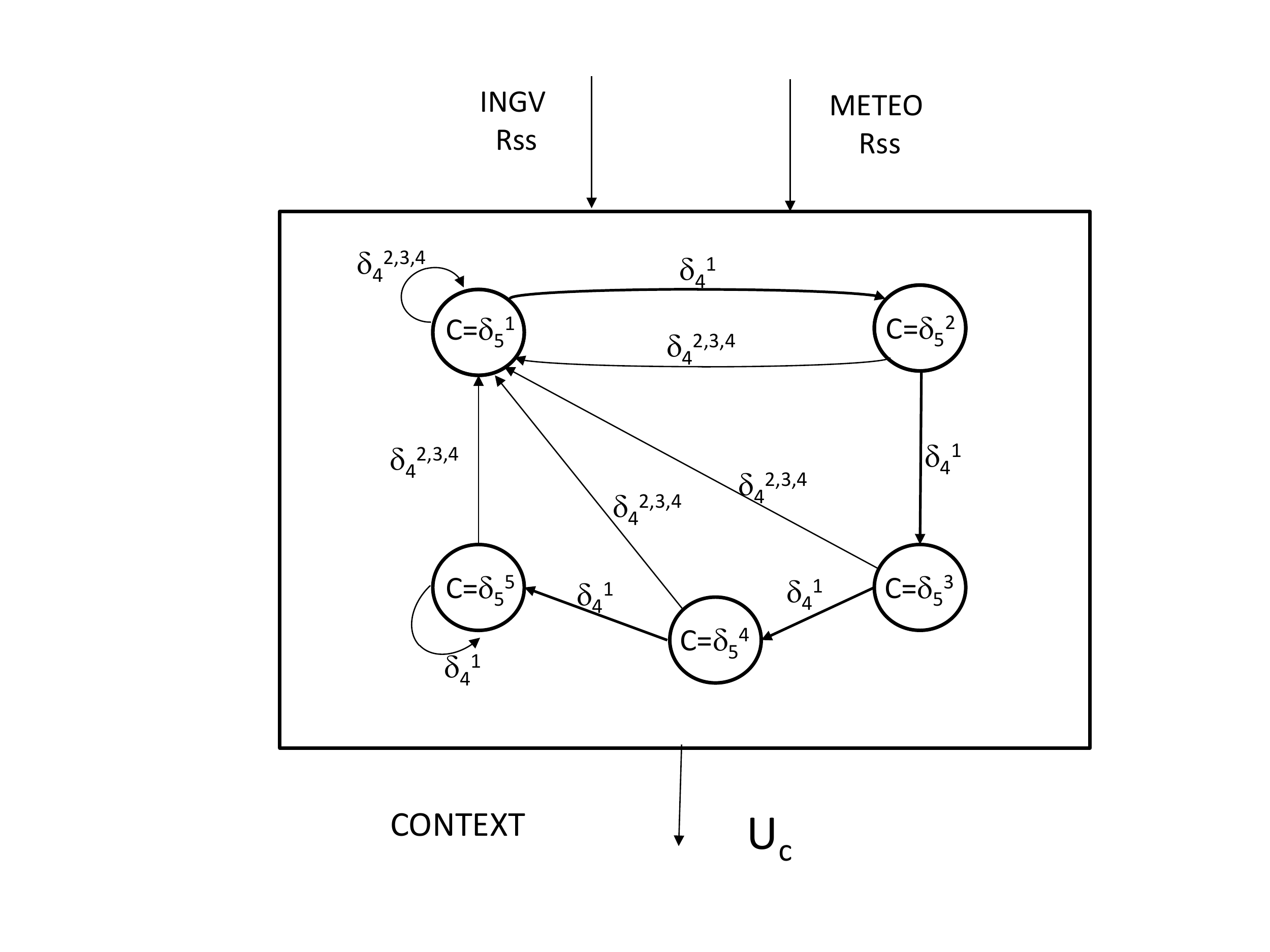}
	\vspace{-1.1 cm}	
	\caption{Context State diagram for the hydrogeological example}
	\label{fig:Rigoctxautoma}
\end{figure}

Figure \ref{fig:Rigoctxautoma} shows the state diagram for the Context automaton.\\

\subsection{The Functional System model}\label{system}

\subsubsection{Functional System Input Variables}

 We assume that, in addition to the CONTEXT-ALERT variable, the Functional System model receives other three input signals from local sensors, so that at the end the input variables determining the system dynamics are the following ones:\\

terrain temperature \{$high, low$\} := V$_{1}$

snow height \{$high, low$\} := V$_{2}$

accelerometer \{$high, low$\} := V$_{3}$

context-alert \{$danger, quiet$\} := V$_{4}$  = U$_c$
\\
\\
\textbf{Input Vector}:

\noindent {\bf v}(t)=
$\left|\begin{array}{cccc}
V_{1}&V_{2}&V_{3}&V_{4}\\V_{1}&V_{2}&V_{3}&\neg V_{4}\\V_{1}&V_{2}&\neg V_{3}&V_{4}\\V_{1}&V_{2}&\neg V_{3}&\neg V_{4}\\
V_{1}&\neg V_{2}&V_{3}&V_{4}\\V_{1}&\neg V_{2}&V_{3}&\neg V_{4}\\V_{1}&\neg V_{2}&\neg V_{3}& V_{4}\\V_{1}&\neg V_{2}&\neg V_{3}&\neg V_{4}\\
\neg V_{1}&V_{2}&V_{3}&V_{4}\\\neg V_{1}&V_{2}&V_{3}&\neg V_{4}\\\neg V_{1}&V_{2}&\neg V_{3}&V_{4}\\\neg V_{1}&V_{2}&\neg V_{3}&\neg V_{4}\\
\neg V_{1}&\neg V_{2}&V_{3}&V_{4}\\\neg V_{1}&\neg V_{2}&V_{3}&\neg V_{4}\\\neg V_{1}&\neg V_{2}&\neg V_{3}&V_{4}\\\neg V_{1}&\neg V_{2}&\neg V_{3}&\neg V_{4}
\end{array}\right|
\in
\Delta_{16}$

\bigskip

\subsubsection{Functional System State}
 
The CONTEXT-ALERT  input is already the result of repeated and consecutive notifications of alert situations, so we may regard it as a variable that is hardly affected by false alarms.  Also, we assume that a disturbance that can instantaneously modify the terrain temperature or the snow height, unless connected with an earthquake, is statistically not very realistic.
On the other hand,  the accelerometer may be a source of false alarms since it can detect a ``high" signal for reasons that are not related to earthquakes: for instance, animals running close to the accelerometer. As a result, we regard as reliable only repeated alerts coming from the accelerometer. So, as in the case of   simultaneous snow and  earthquake warnings, we require, for instance, that the accelerometer has been ``high" for two consecutive time instants (before $t$) in order to regard the information given by the accelerometer as a real warning.
\\ 
We introduce the state variable:\\

ACC-COUNTER\{$0,1, >1$\}
\\

The canonical vector representing the accelerometer counter   is denoted by ${\bf a}$ and  takes values in $\Delta_3=\{\delta^1_{3},\delta^2_3,\delta^3_3\}$.
Specifically, ${\bf a}(t)=\delta^1_3$ if the counter is $0$ at time $t$; 
${\bf a}(t)=\delta^2_3$ if the counter is $1$ at time $t$; and ${\bf a}(t)=\delta^3_3$ if the counter is at least $2$ at time $t$.\\
If the counter at time $t$ has a value in  $\{\delta^1_{3},\delta^2_3\}$ and the  accelerometer vector is ${\bf v}_3(t)=\delta^1_2$, then the counter value at $t+1$ is increased by $1$. If ${\bf a}(t)=\delta^3_3$ and ${\bf v}_3(t)=\delta^1_2$, then 
${\bf a}(t+1)=\delta^3_3$, while when ${\bf v}_3(t)=\delta^2_2$ the counter is moved back to ${\bf a}(t+1)=\delta^1_3$.\\
Therefore, the accelerometer counter updates according to the following   BCN:\\
\\
 ${\bf{a}}(t+1) 
= {\bf{A}}\ltimes {\bf{v}_3}(t)\ltimes {\bf{a}}(t)$, where\\

\noindent $A = \begin{bmatrix} A_1 & A_2  \end{bmatrix}\in {\mathcal L}_{3\times 6}$, and 
\\

\noindent $A_{1}= A \ltimes \delta^{1}_{2} = [\delta^{2}_{3}\ \delta^{3}_{3}\   \delta^{3}_{3}]$\\
$A_{2}= A \ltimes \delta^{2}_{2} = [\delta^{1}_{3}\ \delta^{1}_{3}\ \delta^{1}_{3}]$\\

\subsubsection{Functional System Output}

We assume that the Functional System output can take three values:\\

\noindent ALARM \{$temp-high, snow-high, acc-counter > 1, acc-high,  ctx-danger$\}

\noindent  ATTENTION \{$temp-low,  snow-high, acc-counter-*, acc-*,  ctx-* \ OR\ temp-high, snow-low, acc-counter-*, acc-*, ctx-* \ OR\ temp-high,  snow-high,  acc-counter-low, acc-*, ctx-*\ OR\ temp-high,  snow-high,  acc-counter-*, acc-low, ctx-*\ OR\ temp-high,  snow-high,  acc-counter-*, acc-*,ctx-quiet$\}

\noindent NORMAL \{$temp-low, snow-low, acc-counter-*, acc-*, ctx-*$\} \\

\noindent  Note that the alarm is sent out only when  ``$acc-counter  > 1$" {\em and} ``acc-high". This means that at the time $t^*$ the alarm signal is issued if the accelerometer has detected some movement for at least three consecutive time instants $t^*,\  t^*-1$ and $t^*-2$. Of course, as for the context-alert variable, \emph{the choice of how long we want to wait before issuing the alarm signal is a design parameter that balances conflicting requirements: security on the one hand and the need to avoid false alarms on the other.}

\vspace{0,8 cm}

The functional system  output is denoted by ${\bf m}$ and   takes values in $\Delta_3$.
Based on the previous description of  the three possible output values, it follows that the output vector is generated based on the state ${\bf a}(t)$ and the input ${\bf v}(t)$ according to the following model:\\ 

\noindent ${\bf{m}}(t)  
= {\bf{M}}\ltimes {\bf{v}}(t)\ltimes {\bf{a}}(t)$, where\\

\noindent ${\bf M} = \begin{bmatrix} M_1 & M_2& \dots & M_{16}  \end{bmatrix}\in {\mathcal L}_{6\times 16}$, and 
\\

\noindent  $M_{1}= M \ltimes \delta^{1}_{16} = [ \delta^{2}_{3}\ \delta^{2}_{3}\ \delta^{1}_{3}  ]$\\
$M_{2}= M \ltimes \delta^{2}_{16} = [\delta^{2}_{3}\ \delta^{2}_{3}\ \delta^{2}_{3}  ]$\\
$M_i = M \ltimes \delta^{i}_{16} = M_2, \quad {\rm for}\quad  i=3,\dots, 12$\\
$M_{13}= M \ltimes \delta^{13}_{16} = [\delta^{3}_{3}\ \delta^{3}_{3}\ \delta^{3}_{3} ]$\\
$M_i = M \ltimes \delta^{i}_{16} = M_{13},  \quad {\rm for}\quad  i=14,15, 16$\\

So, overall, the system model is a Boolean Control Network   obtained by connecting the BCN describing the context and the BCN describing the functional model, and hence it is described by the following equations:
\begin{eqnarray}
{\bf{c}}(t+1) &=& {\bf C} \ltimes {\bf u}(t)\ltimes \bf{c}(t) \label{state1}\\
{\bf{a}}(t+1) &=&  {\bf{A}}\ltimes \bf{v}_3(t)\ltimes {\bf{a}}(t) \label{state2}\\
{\bf{v}}_4(t) &=&{\bf H_c}  \ltimes {\bf{c}}(t) \label{cont_out}\\
{\bf{m}}(t) &=& {\bf{M}}\ltimes {\bf{v}}(t)\ltimes {\bf{a}}(t). \label{out}
\end{eqnarray}
Note that the previous system could be represented as a standard BCN having
${\bf u} (t) := {\bf u}(t)\ltimes {\bf v}_1(t)\ltimes {\bf v}_2(t)\ltimes {\bf v}_3(t)$ as input,
${\bf x}(t) := {\bf c}(t)\ltimes {\bf a}(t)$ as state vector, and ${\bf y}(t)={\bf m}(t)$ as output.
Such a representation, however, would be of larger dimension and would not contribute to a better understanding of the system properties. On the contrary, it would make the overall analysis more complicated.  So, 
we investigate the model properties by making use of the previous description 
\eqref{state1}
-\eqref{out}. This provides further evidence of the convenience of using Context-Aware systems to model the system dynamics.
Note that the current cascade structure, having two counter variables as state variables of the two connected BCNs, can be easily adapted to model a large class of Context-Aware  systems that describe a decision process, in particular, an alert system. So, even if we focus on this specific model, it is immediate to understand how the results and properties derived in the following extend to all the alert systems that can be modeled along these same lines.

Let us start by investigating the equilibrium points corresponding to constant inputs.

\section{Equilibria corresponding to constant inputs}\label{equilibria}

Definitions and methods to find equilibrium points   in a system modeled as a BN  have been described in detail in   \cite{BCNCheng,StabBCN2011,EF_MEV_BCN_Aut} to which we refer for further deepening. In this paper we make use of these concepts and characterizations to address equilibrium points of BCNs corresponding to constant inputs. 

\begin{definition}
Given a BCN
\begin{eqnarray}
{\bf x}(t+1) &=& L \ltimes \tilde {\bf u}(t)\ltimes {\bf x}(t), \label{eq_stato}\\
{\bf y}(t) &=& H \ltimes \tilde {\bf u}(t)\ltimes {\bf x}(t), \label{eq_uscita}
\end{eqnarray}
with ${\bf x}(t)\in \Delta_N, \tilde {\bf u}(t)\in \Delta_M$ and ${\bf y}(t)\in \Delta_P$, we say that ${\bf x}_e\in   \Delta_N$ is an {\em equilibrium point} of the BCN corresponding to the constant input 
$\bar {\bf u}$, if 
$$
\left\{
\begin{array}{l}
{\bf x}(0)= {\bf x}_e \cr
\tilde {\bf u}(t)=\bar {\bf u}, \forall\ t\in {\mathbb Z}_+
\end{array}
\right. \qquad \Rightarrow
\qquad
{\bf x}(t)= {\bf x}_e, \forall\ t\in {\mathbb Z}_+.$$
\\
${\bf x}_e\in  \Delta_N$ is a {\em  globally attractive equilibrium point}   of the BCN corresponding to the constant input 
$\bar {\bf u}$, if for every ${\bf x}(0)\in   \Delta_N$ when applying $\tilde {\bf u}(t)=\bar {\bf u}, \forall\ t\in {\mathbb Z}_+$, we obtain that  there exists 
$\tau = \tau({\bf x}(0)) \ge 0$ such that 
${\bf x}(t)= {\bf x}_e, \forall\ t\in {\mathbb Z}_+, t\ge \tau$.\end{definition}
\medskip

Clearly, if  ${\bf x}_e\in  \Delta_N$ is an  equilibrium point of the BCN \eqref{eq_stato}-\eqref{eq_uscita} corresponding to the constant input 
$\bar {\bf u}$, then the corresponding output takes the constant value 
${\bf y}_e := H \ltimes \bar {\bf u}\ltimes {\bf x}_e.$\\
In order to identify the equilibrium points of a BCN corresponding to some constant input $\bar {\bf u}=\delta^k_M$, it is sufficient to evaluate 
the equilibria of the Boolean network  \cite{BCNCheng}
\be
{\bf x}(t+1) = L_k {\bf x}(t).
\label{BNk}
\ee
Such equilibria are the states $\delta^i_N\in  \Delta_N$ such that $\delta^i_N = L_k \delta^i_N$, and hence the states 
$\delta^i_N\in  \Delta_N$ such that $[L_k]_{ii}=1$. Furthermore,   an equilibrium point is   globally attractive (assuming that the input remains constant) if and only if all columns of $L_k^N$ coincide with $\delta^i_N$.\\
In order to identify the equilibrium points of our specific BCN \eqref{state1}
-\eqref{out}, we easily observe that it is sufficient to first identify the equilibria of the context and then those of the functional model.\\
The analysis of \eqref{state1} and the expressions of the matrices ${\bf C}_k, k\in \{1,2,3,4\},$
immediately reveal that 
\begin{itemize}
\item for $\bar {\bf u}=\delta^1_4$ there is a unique equilibrium point ${\bf c}_e= \delta^5_5$, and it is  globally attractive;
\item for $\bar {\bf u}=\delta^i_4, i\ne 1,$ there is a unique equilibrium point ${\bf c}_e= \delta^1_5$, and it is  globally attractive, in turn.
\end{itemize}
The constant output value corresponding to the two cases are $\bar {\bf v}_4= \delta^1_2$ for $\bar {\bf u}=\delta^1_4$, and 
$\bar {\bf v}_4= \delta^2_2$ for $\bar {\bf u}=\delta^i_4, i\ne 1.$
\\
Let us now consider the functional model \eqref{state2}. We have the following two cases:
\begin{itemize}
\item for $\bar {\bf v}_3=\delta^1_2$ there is a unique equilibrium point ${\bf a}_e= \delta^3_3$, and it is globally attractive;
\item for $\bar {\bf v}_3=\delta^2_2$ there is a unique equilibrium point ${\bf a}_e= \delta^1_3$, and it is globally attractive, in turn.
\end{itemize}
So, to summarize, we have the following situation:
\bigskip

\begin{center}
\begin{tabular}{c|c}
Constant input & Equilibria \\
\hline
$\bar {\bf u} = \delta^1_4, \bar {\bf v}_3 = \delta^1_2$& ${\bf c}_e= \delta^5_5$, ${\bf a}_e= \delta^3_3$\\
$\bar {\bf u} = \delta^1_4, \bar {\bf v}_3 = \delta^2_2$& ${\bf c}_e= \delta^5_5$, ${\bf a}_e= \delta^1_3$\\
$\bar {\bf u} = \delta^i_4, i\ne 1, \bar {\bf v}_3 = \delta^1_2$ & ${\bf c}_e= \delta^1_5$, ${\bf a}_e= \delta^3_3$\\
$\bar {\bf u} = \delta^i_4, i\ne 1, \bar {\bf v}_3 = \delta^2_2$& ${\bf c}_e= \delta^1_5$, ${\bf a}_e= \delta^1_3$
\end{tabular}
\end{center}
\medskip

If we now introduce the remaining inputs, and we recall that the  input $\bar {\bf v}_4$ is the  output of the context model, we obtain the following results that describe for each constant input the equilibria and the corresponding constant outputs.
\bigskip

\begin{center}
\begin{tabular}{c|c|c}
Constant input & Equilibria & Constant output \\
\hline\\
$\begin{array}{c}\bar {\bf v}_1=\delta^1_2, \bar {\bf v}_2=\delta^1_2,\cr  \bar {\bf v}_3 = \delta^1_2, \bar {\bf u} = \delta^1_4\end{array}$& ${\bf c}_e= \delta^5_5$, ${\bf a}_e= \delta^3_3$ & ${\bf m}_e= \delta^1_3$\\\hline\\
$\begin{array}{c}\bar {\bf v}_1=\delta^2_2, \bar {\bf v}_2=\delta^2_2 \cr
 \bar {\bf v}_3, \bar {\bf u} \ {\rm arbitrary }\end{array}$&$\begin{array}{c} ({\bf c}_e, {\bf a}_e)\in \cr \{
 (\delta^1_5, \delta^1_3),  (\delta^1_5, \delta^3_3), (\delta^5_5, \delta^1_3), (\delta^5_5, \delta^3_3)\}\end{array}$
 & ${\bf m}_e= \delta^3_3$\\\hline\\
 {\rm all other choices} &$\begin{array}{c} ({\bf c}_e, {\bf a}_e)\in\cr  \{
 (\delta^1_5, \delta^1_3),  (\delta^1_5, \delta^3_3), (\delta^5_5, \delta^1_3), (\delta^5_5, \delta^3_3)\}\end{array}$
 & ${\bf m}_e= \delta^2_3$
\end{tabular}
\end{center}
\medskip

This analysis shows that  
\emph{ no limit cycles can appear, and hence no contradicting alarm messages can be delivered by the system}.

\section{Observability and reconstructibility}\label{observability}

 The definitions of observability and reconstructibility are given by slightly adjusting those given in  \cite{EF_MEV_BCN_obs2012}, since in this   paper we assume that the input at time $t$ directly affects the update of the output at time $t$ (we consider  proper BCNs as opposed to the strictly proper BCNs, typically investigated in the literature). These properties have been the subject of an extensive research, in particular we refer the interested reader to \cite{MM_obs,Zhang_obs}.

\begin{definition}
Given a BCN \eqref{eq_stato}-\eqref{eq_uscita}, with ${\bf x}(t)\in \Delta_N, \tilde {\bf u}(t)\in \Delta_M$ and ${\bf y}(t)\in \Delta_P$, we say that 
the BCN is
\begin{itemize}
\item {\em observable} if there exists $T\in {\mathbb Z}_+$ such that the knowledge of the input and output vectors in the discrete interval $\{0,1,\dots, T\}$ allows to uniquely determine the initial state ${\bf x}(0)$;
\item {\em reconstructible} if there exists $T\in {\mathbb Z}_+$ such that the knowledge of the input and output vectors in the discrete interval $\{0,1,\dots, T\}$ allows to uniquely determine the final state ${\bf x}(T)$.
\end{itemize}
\end{definition}
\medskip

The hydrogeological model proposed in this paper is not observable.
Indeed, it is easily seen that the first BCN:
\be
\begin{array}{rcl}
{\bf{c}}(t+1) &=& {\bf C} \ltimes {\bf u}(t)\ltimes \bf{c}(t) \cr
{\bf{v}}_4(t) &=&{\bf H_c}  \ltimes {\bf{c}}(t) 
\end{array}
\label{primaBCN}
\ee
is not observable, since initial
 states as ${\bf c}(0)=\delta^1_5$ and ${\bf c}(0)=\delta^2_5$ corresponding to the constant input $\tilde {\bf u}(t)=\delta^4_4, t\in {\mathbb Z}_+$, 
 generate the same output sequence ${\bf{v}}_4(t) = \delta^2_2, t\in {\mathbb Z}_+$.
 This clearly prevents the whole interconnected BCN representing the hydrogeological system to be observable.
 It is worth remarking, however, that the system is observable in a weak sense (see  \cite{MM_obs,Zhang_obs}), since for every pair of initial states there exists a specific choice of the input sequence that would generate  two distinct output trajectories from which the initial states could be recognised.
 \\
 
 Lack of observability is not a major issue. In particular, observability does not seem to be a fundamental system property 
 for the hydrogeological model, since identifying the initial state of the system during some observation interval does not bring any practical advantage.
 On the other hand, reconstructibility is a more relevant property to investigate:  by identifying the current system state, say ${\bf x}(T)$, from the observation of  the input and the output in some time interval $[0,T]$, \emph{one may anticipate whether an alert signal will lead to an alarm signal at the next time instant or not and hence be ready to run away or to provide support}.

\begin{proposition} The hydrogeological system described by
\eqref{state1}-\eqref{state2}-\eqref{cont_out}-\eqref{out}
is reconstructible and   the definition of reconstructibility holds for $T=4.$
\end{proposition}
 
{\sc Proof.}
 We exploit the fact that the overall system is represented as a cascade of two BCNs 
 and first investigate the possibility of identifying the context state from the available information.
 We note that every time the input ${\bf u}(t)$ is equal to $\delta^1_4$ the counter variable ${\bf c}$ increases when moving from $t$ to $t+1$, unless it has already reached the maximum value in which case it remains constant to the value ${\bf c}(t+1)={\bf c}(t)=\delta^5_5$.
 Conversely, if  ${\bf u}(t)\ne \delta^1_4$ then, independently of ${\bf c}(t)$, the counter state ${\bf c}(t+1)$ takes the value $\delta^1_5$.
 This implies that if we know the input sequence ${\bf u}(t), t\in {\mathbb Z}_+$, (even if we ignore the output) then at latest at $T_1=4$ 
 (a situation that occurs only if ${\bf c}(0)=\delta^1_5$ and ${\bf u}(t)=\delta^1_4$ for $t=0,1,2,3$) 
 we are able to identify exactly ${\bf c}(T_1)$. Therefore the state of the context is always reconstructible (indeed, based on the input sequence alone).
 If we identify ${\bf c}(T_1)$ then, knowing ${\bf u}(t)$ for $t\ge T_1$, we are able to determine ${\bf c}(t)$ 
 and ${\bf v}_4(t)$ 
 for $t\ge T_1$.
 \\
 At the same time, we now know ${\bf v}(t), t\ge T_1$, which is the input of the Functional System model:
 \be
 \begin{array}{rcl}
{\bf{a}}(t+1) &=&  {\bf{A}}\ltimes \bf{v}_3(t)\ltimes {\bf{a}}(t) \cr
{\bf{m}}(t) &=& {\bf{M}}\ltimes {\bf{v}}(t)\ltimes {\bf{a}}(t). \end{array}
\label{secondaBCN}
\ee
So, if we prove the reconstructibility of this second system, we have proved the reconstructibility of the overall 
hydrogeological system.
Reconstructibility of \eqref{secondaBCN} is easily proved along the same lines as for the first BCN. 
Indeed,   every time the input ${\bf v}_3(t)$ is equal to $\delta^1_2$ the counter variable ${\bf a}$ increases when moving from $t$ to $t+1$,
unless it has already reached the maximum value in which case  it remains constant to the value  ${\bf a}(t+1)={\bf a}(t)=\delta^3_3$.
 Conversely, if  ${\bf v}_3(t)\ne \delta^1_2$ then, independently of ${\bf a}(t)$, ${\bf a}(t+1)$ takes the value $\delta^1_3$.
 This means that if we know the input sequence ${\bf v}_3(t), t\in {\mathbb Z}_+$, then at latest at $T_2=2$ 
 we are able to identify exactly ${\bf a}(T_2)$ (and hence ${\bf a}(t), t\ge T_2$). \\
 Putting together the two parts of the reasoning, we can claim that from $T=\max\{T_1,T_2\}=4$ onward, we are able to identify both ${\bf c}(t)$ and ${\bf a}(t)$. This proves that the system is reconstructible and the definition holds for $T=4$.
\hfill$\clubsuit$
 \medskip

\section{Fault detection}\label{fault}

A general theory of fault detection in the context of BCNs is still at an early stage, nonetheless there have been some interesting contributions 
addressing this important problem \cite{FaultDetTAC15,FaultBCN_CCC2015,arXiv2019FaultBCN,ZhangFaultACC2018}. 
The fault detection problem investigated in  \cite{FaultDetTAC15,FaultBCN_CCC2015,ZhangFaultACC2018} refers to the case when the matrices $L$ and $H$ involved in the state and output equations \eqref{eq_stato} and
\eqref{eq_uscita} are replaced by two different (and potentially arbitrary) logical matrices $L_F$ and $H_F$, as a consequence of a fault.
In the context of the hydrological system (and of any alarm system for which the  alarm is launched only when some variable take critical values on a sufficiently long time interval), the state variables represent counter variables, and the state-update equations are extremely elementary. So, the case of a fault that arbitrarily affects the matrices that regulate the counter variables update  does not seem a very realistic one.
Similarly, the case when the matrix that generates the alarm/alert output signal is replaced by a different logic matrix seems too general and not representative of the real faults that may affect the system. 
 
An exception is represented by the case when the change of the matrices $L$ or $H$ formalizes a very classical type of fault that has been investigated for logic circuits: the so called {\em stuck-in fault}. 
 In the context of the hydrogeological system, this corresponds to the case when 
one (or more) state variable is stuck at some constant value, independently of the soliciting input. In other words, we are considering the case when one of the counters for some reasons does not update (the case when both counters get stuck  at the same time is extremely unrealistic).


Assume that there exists some time instant $\tau\in {\mathbb Z}_+$  such that either ${\bf c}(t)= {\bf c}(\tau)$ for every $t\ge \tau$ or ${\bf a}(t)={\bf a}(\tau)$ for every $t\ge \tau$. The problem that we want to investigate is the following one: Assuming that we know the hydrogeological system model \eqref{state1}-\eqref{state2}-\eqref{cont_out}-\eqref{out}, and that we have access both to the inputs  ${\bf u}(t), {\bf v}_1(t), {\bf v}_2(t), {\bf v}_3(t)$  and to the output
 ${\bf y}(t)={\bf m}(t)$,
are we able to detect such a fault? If so, are we able to identify  which of the two counters is blocked?
\medskip

The observability and reconstructibility analysis carried on in the previous section allows  to provide a quite complete answer   to both  questions. 
It is in fact easy to see that if the stuck-in fault affects the context state variable at $t=\tau$ and blocks the context state at the value ${\bf c}(t)= \delta^i_5, i\ne 5,$ for every $t\ge \tau$, then every input signal ${\bf u}(t), t\in {\mathbb Z}_+,$  that starting from $t= \tau$ does not take the value $\delta^1_2$ for more than 3 consecutive time instants, will generate the constant context output   signal ${\bf v}_4(t) = \delta^2_2$ (context-alert= quiet),
exactly as it would if the context state would be correctly working starting from that same value 
${\bf c}(t)= \delta^i_5$ at $t=\tau$ but correctly evolving in time. This is due to the fact that the system is not observable and hence different state trajectories are compatible with the same input-output trajectories; in particular, there exist constant state trajectories that cannot be distinguished from time-varying  state trajectories.
A quite similar analysis could be carried on for the Functional System state variable ${\bf a}(t)$, since the case when ${\bf a}$ is stuck at $\delta^1_3$ or at $\delta^2_3$ cannot be detected from the output signal ${\bf m}(t)$, when ${\bf v}_3$ is identically equal to $\delta^2_2$. This allows to say that, in general, a stuck-in fault may not be detected and hence, a fortiori, identified.\\
This is surely not a good system feature. However, it must be remarked that the situations we have depicted are those 
when an alarm signal would have not been generated even if the system state would have not been stuck at a constant value. Indeed if the input signals ${\bf u}(t), {\bf v}_1(t), {\bf v}_2(t), {\bf v}_3(t)$ simultaneously take the value $\delta^1_2$ on a sufficiently large time window, then stuck-in faults that would erroneously lead to a non-alarm signal could be easily identified and hence corrected.

Indeed, the simple knowledge of ${\bf u}(t), t\in {\mathbb Z}_+,$ allows to identify ${\bf c}(t)$ for $t\in {\mathbb Z}_+, t\ge T_1=4$. Similarly, the knowledge of ${\bf v}_3(t), t\in {\mathbb Z}_+,$ allows to identify ${\bf a}(t)$ for $t\in {\mathbb Z}_+, t\ge T_2=2$.
This means that we can obtain an estimate $\hat {\bf c}(t)$ of ${\bf c}(t)$ and an estimate  $\hat {\bf a}(t)$ of ${\bf a}(t)$, and these estimates are both exact from $T=T_1=4$ onward, provided that the system is not affected by faults.
By making use of these estimates and  the system model, we can derive the estimate of the 
context output
\be
 \hat {\bf{v}}_4(t) = {\bf H_c}  \ltimes \hat {\bf{c}}(t) \label{est_cont_out}
\ee
and of the overall system
output  
\be
\hat {\bf{m}}(t) = {\bf{M}}\ltimes \hat {\bf{v}}(t)\ltimes \hat {\bf{a}}(t), \label{est_out}
\ee
where $\hat {\bf{v}}(t) := {\bf v}_1(t)\ltimes {\bf v}_2(t) \ltimes {\bf v}_3(t)\ltimes \hat {\bf v}_4(t).$
\\
So, if the inputs are all equal to $\delta^1_2$ for at least 4 consecutive time instants, we know that an alarm signal should be generated and if this is not the case then a fault has necessarily occurred.\\

To conclude, we have proved what follows for the hydrogeological model:

\begin{proposition}  Given the hydrogeological system described by
\eqref{state1}-\eqref{state2}-\eqref{cont_out}-\eqref{out}, a stuck-in fault for one of the state variables,  ${\bf c}(t)$ or ${\bf a}(t)$, cannot   be identified corresponding to all the  input sequences, but if one of the counters gets stuck at a value that is not maximum, thus preventing the possible generation of an alarm, then the previous state estimator always allows to detect and identify the stuck-in fault at latest after  $T=4$ times instants after the fault has occurred.
\end{proposition}

Note, finally, that  a false alarm cannot possibly be issued, because this would require not only that one of the counters is stuck to the maximum value but also that the other is at the maximum value in turn and  the inputs are all high, but \emph{this is the case when the alarm message should be issued!} 
  \bigskip
  
\section{Conclusions and future work}\label{conclusion}

  In this paper we model a simple Context-aware system as a Boolean Control Network in order to use the powerful tools typical of system theory, which apply to  linear analog systems, also to digital systems, whose  properties are usually expressed by logical rules. The ultimate goal is to pave the way to formally assess reliability and safety properties of self adapting  safety critical systems. \\
The existence of globally attractive equilibrium points under constant input and the reconstructibility of the system have been proved, as well as the possibility of identifying some faults which could adversely affect the system output.\\
Further work is to be made to apply these techniques to more complex feedback systems, where the output of the functional systems can affect in turn the state of the Context, and to enhance fault tolerance by considering possible correlations among the sensors and other system input/output devices.

\end{document}